          [2001/04/25 1.1 (PWD)]
\documentclass[a4paper,twocolumn]{esapub2005} 
\usepackage{times}
\usepackage{natbib}
\usepackage{epsfig}

\title{Evidence of an association between the presence of penumbrae and strong radial outflows in sunspots}
\author{S. Vargas Dom\'inguez}
\author{J. A. Bonet}
\author{V. Mart\'inez Pillet}
\affil{Instituto de Astrof\'isica de Canarias, Tenerife-Spain, E-mail: svargas@iac.es}
\author{Y. Katsukawa}
\affil{National Solar Observatory, Tokio-Japan}

\begin{document}

\maketitle


\begin{abstract}
Time series of high-resolution images of the complex active region NOAA 10786 are studied. The observations were performed in G-band (430.5 nm) and in the nearby continuum (463.3 nm), on July 9, 2005 at the Swedish 1-meter Solar Telecope (SST) in La Palma. Granular proper motions in the surroundings of the sunspots have been quantified. A large-scale radial outflow in the velocity range 0.3 - 1 $km$ $s^{-1}$ has been measured around the sunspots by using local correlation tracking techniques. However, this outflow is not found in those regions around the sunspots with no penumbral structure. This result evidences an association between penumbrae and the existence of strong horizontal outflows (the moat) in sunspots.
\end{abstract}

\section{Observations}
\begin{figure}
\centering
\begin{tabular}{c}
\includegraphics[width=0.81\linewidth]{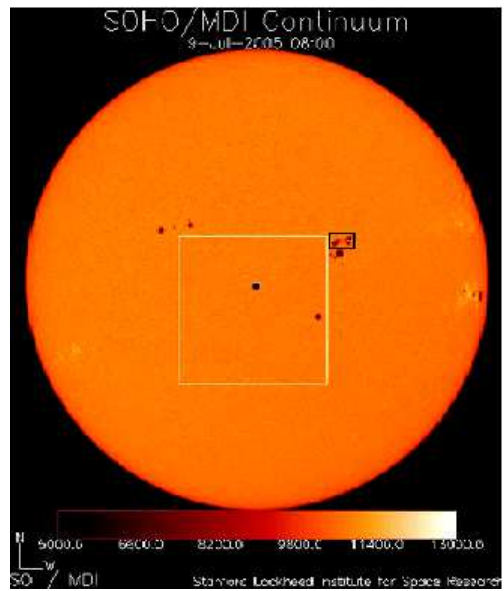}
\end{tabular}
\caption{Solar disc with the active region NOAA 10786 (black box) observed on July 9, 2005 (Courtesy SOHO).
\label{observations}}
\end{figure}

The active region NOAA 10786 was observed on July 9, 2005 at the Swedish 1-meter Solar Telescope, La Palma  \citep{scharmer03} during the International Time Program. This complex region observed at heliocentric position $\mu=0.9$ (Figure~ \ref{observations}), shows a $\delta$-configuration (Figure~\ref{images}).\\

Two simultaneous time sequences of high resolution \mbox{images} were taken in $430.5 \pm 0.54$ nm (G-band) and $436.6 \pm 0.57$ nm (G-cont), respectively. The sequences span for 79 minutes, from 7:47 UT to 9:06 UT and the pixel size is $0".041$. The observational strategy consisted in taking G-band images and simultaneous G-cont phase diversity image-pairs required for the restoration technique (i.e. three different channels.)

\section{Data Processing}
The Multi-Object Multi-Frame Blind Deconvolution (MOMFBD) technique  \citep{noort05}, an extension of the phase diversity one but for multiple objects, has been applied to restore the images.

\begin{figure*}
\centering
\begin{tabular}{cc}
\includegraphics[width=0.68\linewidth]{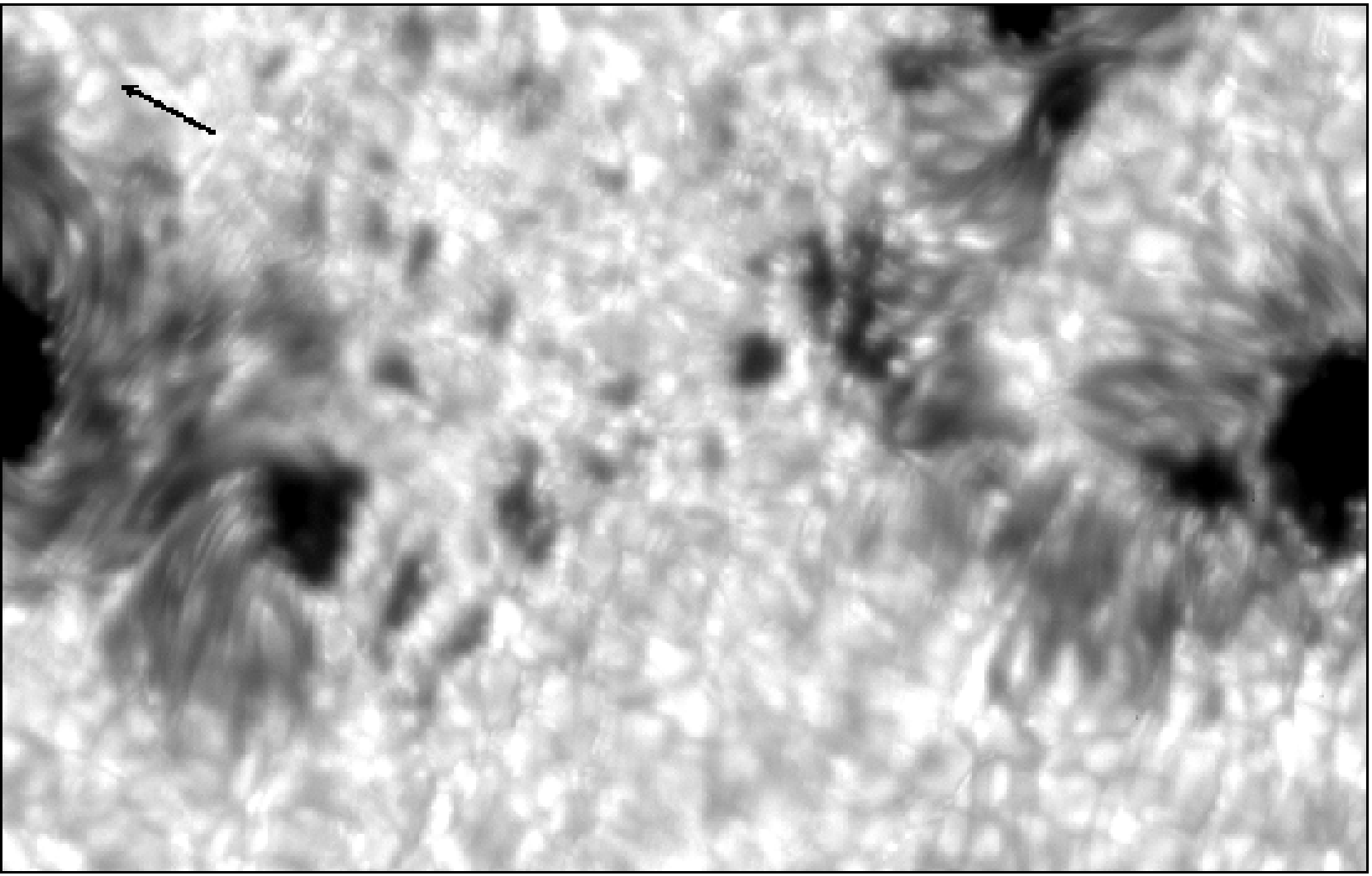}\\
\includegraphics[width=0.68\linewidth]{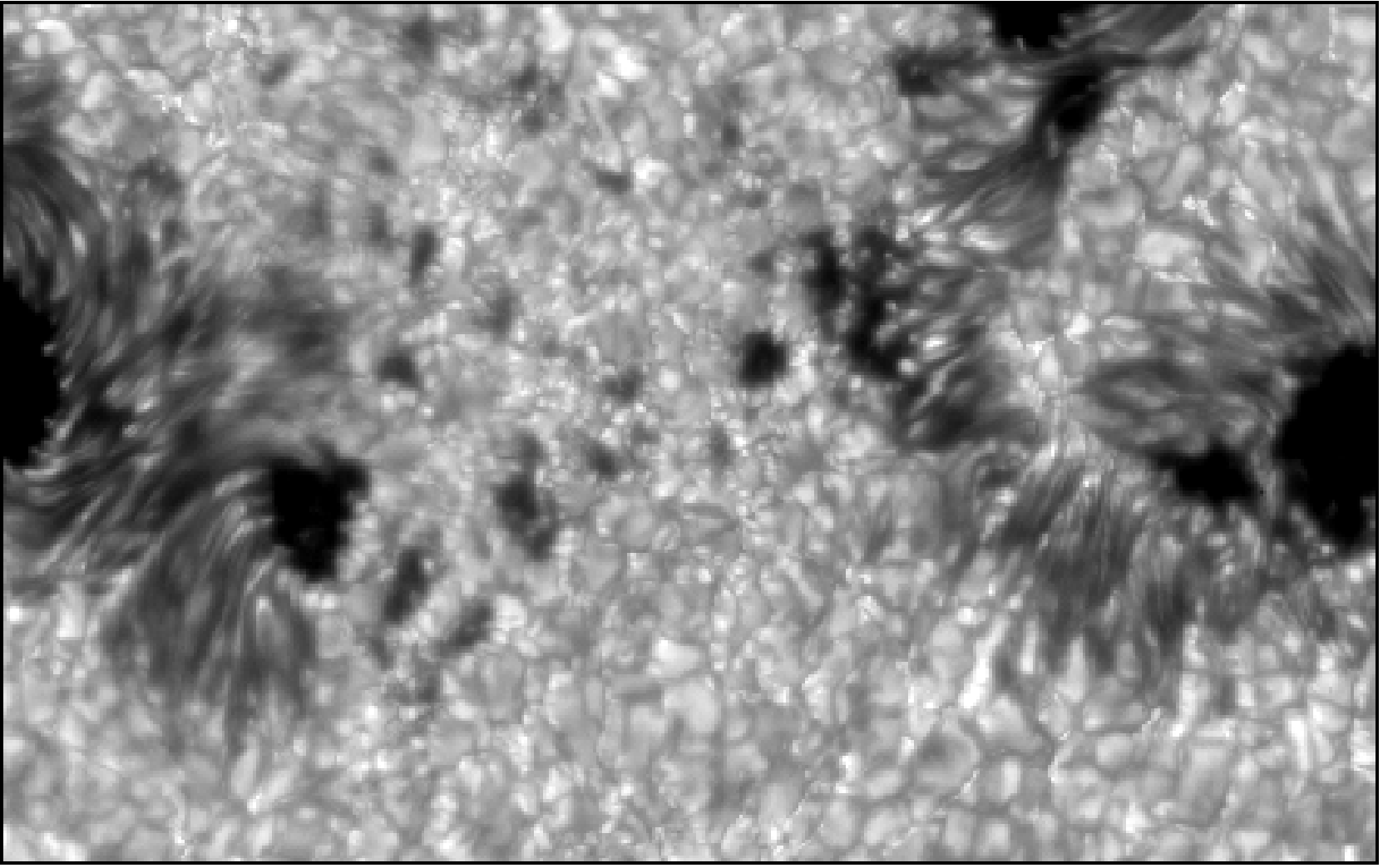}\\
\includegraphics[width=0.68\linewidth]{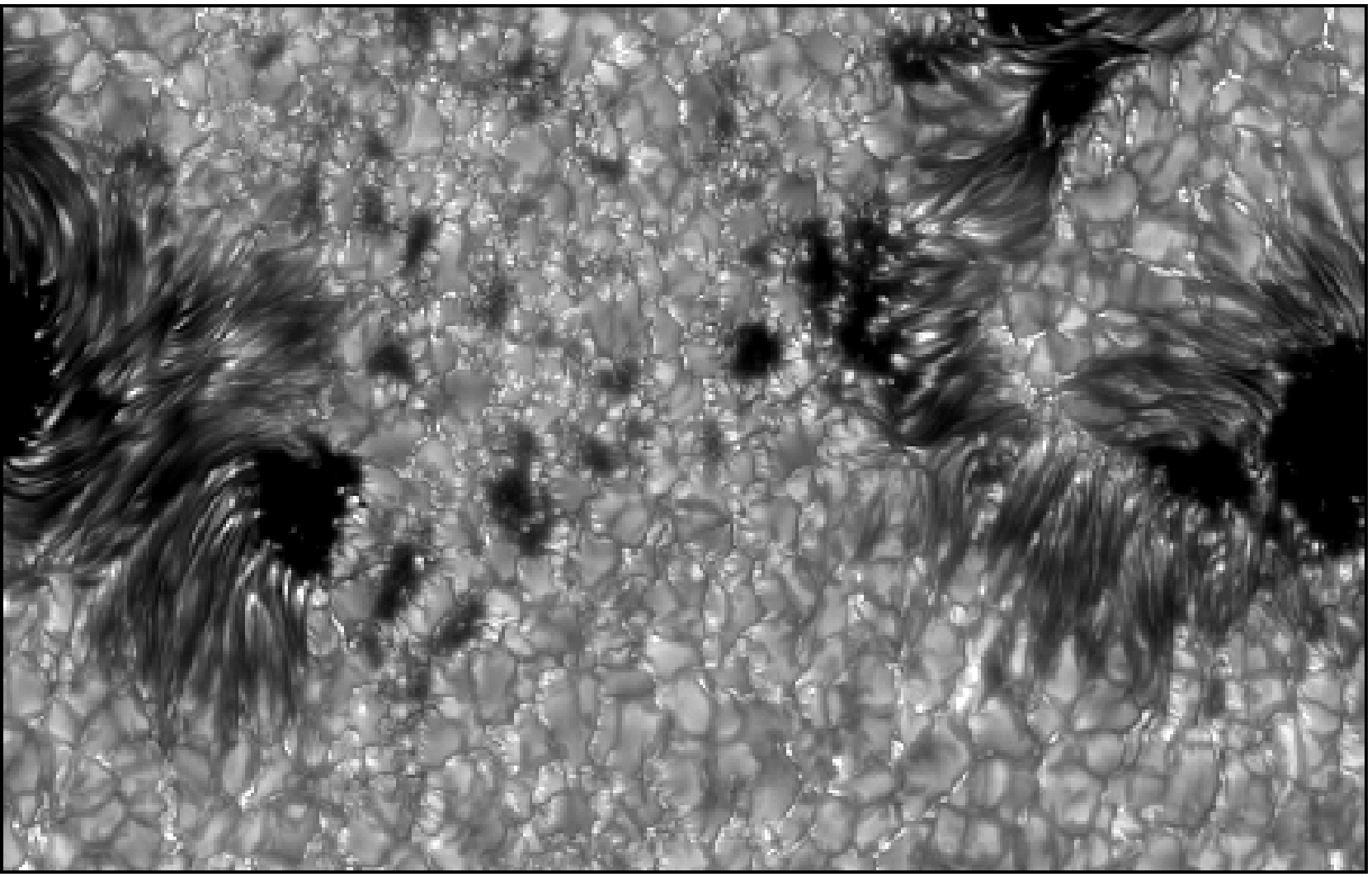}
\end{tabular}
\caption{Sunspots images from the 430.5 nm (G-band) series. Images of poorest quality (upper pannel) and best quality (middle pannel) within a G-band 18 images-set used for restoration. The lower pannel shows the resulting restored image. The arrow in the upper pannel points to the solar disc centre and the final FOV after restorations is 57".8 $\times$ 34".4
\label{images}}
\end{figure*}

Sets of about 18 images per channel (i.e. $3 \times 18$ images) have been combined to produce pairs of simultaneous G-band and G-cont restored images. Each resulting time series, consists of 472 restored images with a cadence of 10.0517 s. Figure~\ref{images} shows an example of restoration in G-band for one of the sets.

Finally, the images have been corrected for diurnal field rotation, rigidly aligned, destretched and subsonic filtered with a cut-off phase velocity of 4 $km$ $s^{-1}$ \citep{title89} to produce the series for further analysis.\\

\section{Analysis and results}

\subsection{Motions of granules}

Proper motions of the structures in the field-of-view (FOV) have been studied by local correlation tracking techniques (LCT, \citep{november88}). A number of tracking window sizes and averaging time windows have been tried.\\

\begin{figure}
\centering
\includegraphics[width=1.\linewidth]{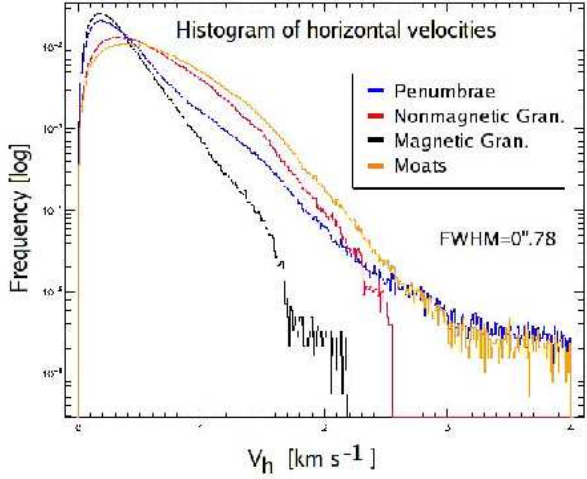}
\caption{Histogram of horizontal velocities $v_h$ (5 \mbox{minutes} average; FWHM=0".78) derived from a local correlation tracking technique. \label{histogram}}
\end{figure}

The histogram of horizontal velocities (5 min average; Gaussian tracking window FWHM=0".78) in Figure~\ref{histogram}, shows the contributions from different regions within the FOV (see also Table~\ref{tablevel}). The highest proper motions up to 3 $km$ $s^{-1}$ are mainly observed in the penumbrae.\

\begin{table}
\begin{center}
\caption{Statisticts of the horizontal velocities [$km$ $s^{-1}$] at various regions within the FOV, for two different tracking windows.}\vspace{1em}
    \renewcommand{\arraystretch}{1.2}
     \begin{tabular}[h]{rrrrrrrrrr}
     &&&&&& ~ ~ ~ ~ \small{FWHM=0."2}     &  \small{FWHM=0."78} & \\
     \end{tabular}
     \begin{tabular}[h]{lcccc}
      REGION & rms & mean & rms & mean \\\hline 
      Penumbrae  & 0.56 & 0.76 & 0.32 & 0.39 \\
      Nonmagnetic Gran.  & 0.66 & 1.00 & 0.36 & 0.56  \\
      Magnetic  Gran.  & 0.54 & 0.66 & 0.22 & 0.31 \\
      Moats  & 0.71 & 1.06 & 0.40 & 0.64  \\
      \hline \\
      \end{tabular}
    \label{tablevel}
  \end{center}
\end{table}

\begin{figure*}
\centering
\includegraphics[width=.8\linewidth]{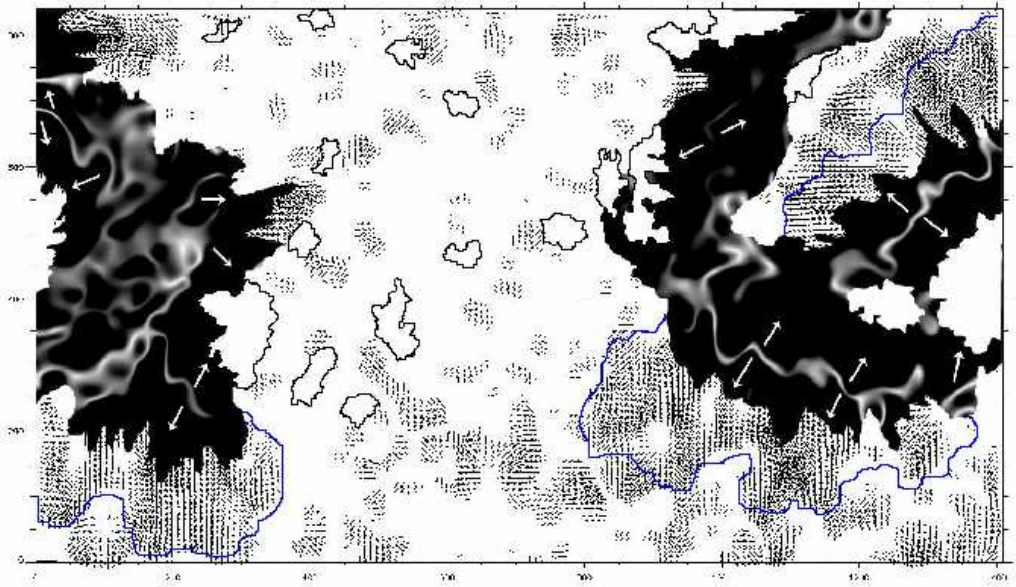}
\caption{Flow map (FWHM=0".78) showing the horizontal velocities in a range between 0.3 and 1 $km$ $s^{-1}$. The blue contours outline the strong radial outflows (moats) surrounding the sunspots. These moats are found to be closely associated to the existence of penumbrae.
Penumbral regions are represented by black areas. Bright filamentary structures inside penumbrae correspond to small horizontal velocities ($v_h < 0.13$ $km$ $s^{-1}$) and mark the division between two tendencies: flows toward the umbra in the inner penumbra and toward the surrounding photosphere in the outer penumbra (see white arrows in the figure).
The lenght of the black bar at coordinates (0,0) corresponds to 4 $km$ $s^{-1}$. The coordinate unit is 0".041 (1 pixel).
\label{moats}}
\end{figure*}

Figure~\ref{moats} shows the flow map (70 min average, FWHM=0".78) for velocities ranging from 0.3 to 1 $km$ $s^{-1}$. Strong radial outflows (moats) are evident surrounding the sunspots. We have found that these moats are closely associated with the existence of penumbra and are missing in the sides with no penumbral structure.\\

Following \citep{november87}  (see also \citep{marquez06}) we have calculated from the horizontal velocities, the divergence structures and subsequently the vertical velocities by using the equation (\ref{eq:November}).

\begin{equation}
v_z=h_m \nabla.v(x,y)
\label{eq:November}
\end{equation}

where $h_m=150$ $km$ stands for the scale height of the flux of mass, and $\nabla.v(x,y)$ is the divergence of the horizontal velocity field.\\

\begin{figure*}
\centering
\includegraphics[width=0.8\linewidth]{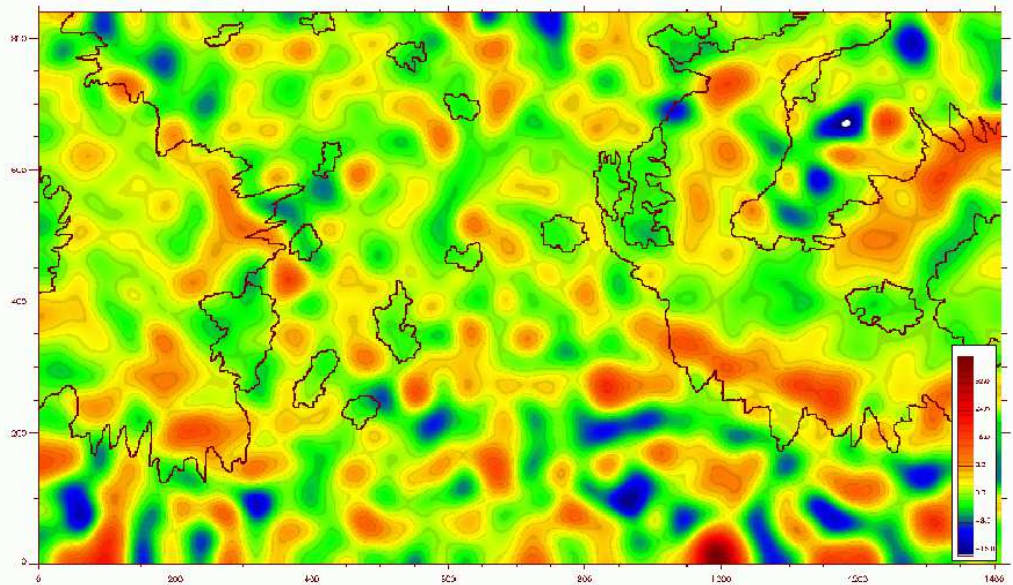}
\caption{Map of vertical velocities  (70 minutes average, FWHM=0".78). Extreme values of divergence can be easily identified: exploding granules at coordinates (1000,0) and (825,275), and powerful sinks at the upper right part of the image. The coordinate unit is 0".041 (1 pixel).
\label{divergen}}
\end{figure*}

Figure~\ref{divergen} shows the resulting vertical flows in the FOV, ranging from -20 $m$ $s^{-1}$ (downward) to 40 $m$ $s^{-1}$ (upward). The extreme values of vertical velocities (see the caption in Figure\ref{divergen}) correspond to strong exploding granules and powerful sinks, respectively.\\

\subsection{Motions of centres of divergence}

The divergence structures displace in time dragged by flows at scales larger than the mesogranular ones. We aim at determine the map of such large-scale flows. Following  the procedure explained above, and averanging over 5 min windows, a sequence of 14 maps of divergence structures was obtained during a 70-min period within the series. The application of LCT techniques (but now with a tracking window of  FWHM=3") to the structures in the sequence of 14 maps of divergence, allows us to produce an average velocity map for the large-scale flows. \\

\begin{figure*}
\centering
\includegraphics[width=.8\linewidth]{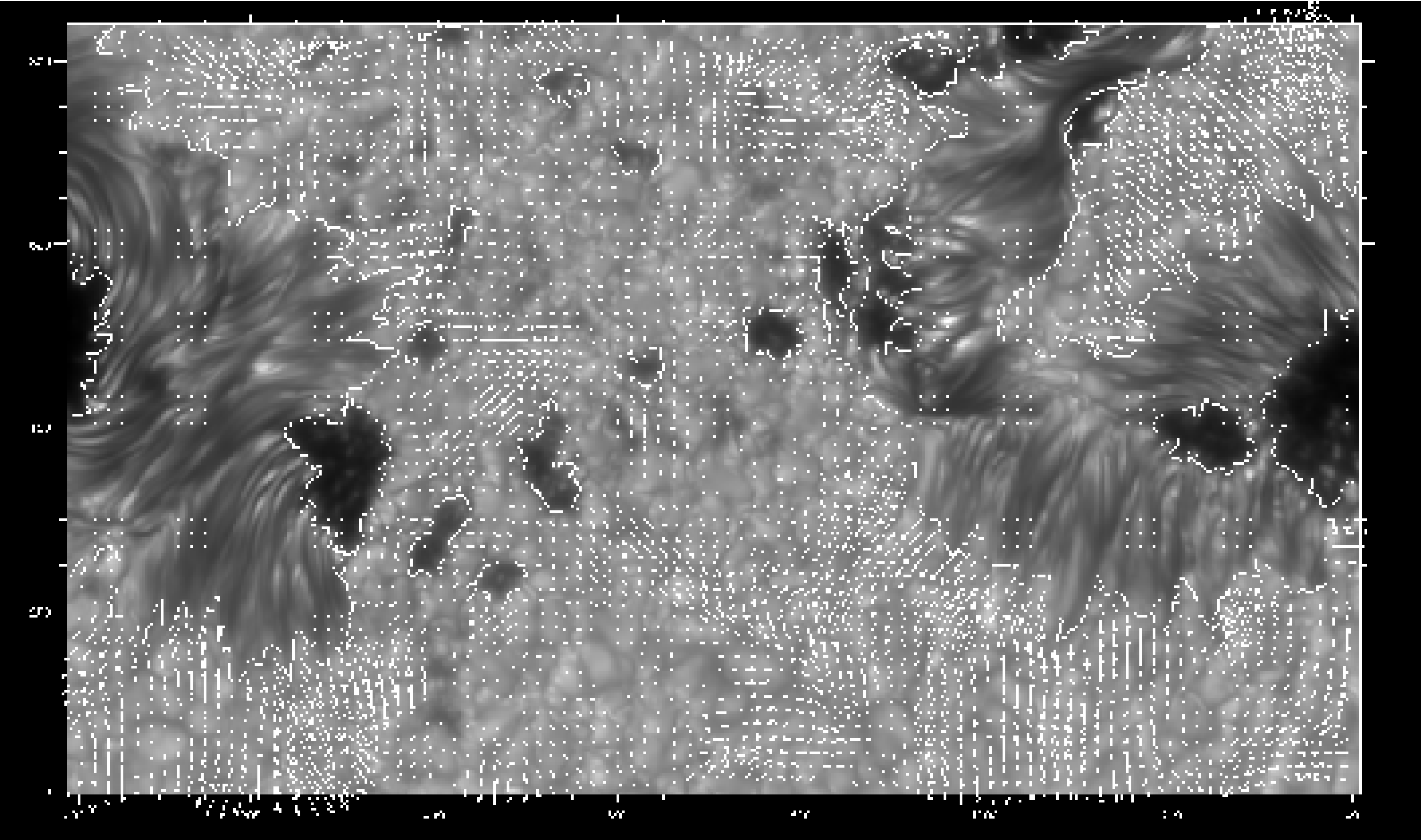}
\caption{Centres of divergence displace in time as showed in this flow map. The moats are clearly identified as rapid outflows regions as in Figure~\ref{moats}. The velocity range for the arrows goes from 0 to 1.18 $km$ $s ^{-1}$. The coordinate unit is 0".041 (1 pixel).
\label{flowdiv}}
\end{figure*}

The resulting flow map for the divergence structures (Fg.~\ref{flowdiv}) looks similar, although much less noisy, to the one obtained when using granules as tracers. Now the moat reveals even more clearly outlined around the sunspots, but again only in those regions with penumbra.\\

\section{Discussion and conclusions}

Time series of broad band images for a complex $\delta$-configuration
sunspot spanning over 71 minutes were restored for atmospheric and instrumental degradation. We were able to study motions of granules and mesogranules surrounding the sunspots and derived velocity ranges corresponding to the so-called moat between 0.3 and 1 $km$ $s^{-1}$.\\

The main result from our analysis is the fact that the moat, defined as strong radial outflows from the sunspots, is only found surrounding those parts of the sunspots where the penumbra exists. This result has been found not only by computing the horizontal flows using the granules as tracers, but also when identifying the large-scale outflows surrounding the sunspots by the displacements of the centres of divergence. These large-scale outflows in Figure \ref{flowdiv} are more uniform and better organized than the horizontal ones in Figure \ref{moats}.\\

We have also identified inside the penumbrae, the divisory lines which separate the flows goind toward the umbra in the inner penumbrae and toward the surrounding photosphere in the outer penumbra. Figure \ref{moats} shows these lines as filamentary bright structures corresponding to small horizontal velocities ($v_h < 0.13$ $km$ $s^{-1}$).

\section*{Acknowledgments}

The authors are grateful to Mats Lofdahl, Michiel Van Noort and Luc Rouppe van der Voort, for their inputs about the restoration process. The Swedish 1-m Solar Telescope is operated on the island of La Palma by the Institute of Solar Physics of the Royal Swedish Academy of Sciences in the Spanish Observatorio del Roque de los Muchachos of the Instituto de Astrof\'isica de Canarias. Partial support by the Spanish Ministerio de Educaci\'on y Ciencia through project ESP2003-07735 is gratefuly \mbox{acknowledged}.


\end{document}